%% file: main.tex
\documentclass[twocolumn,onecolappendix]{aastex7}

\input{./preamble.tex}

\received{December 23, 2025}
\revised{March 10, 2026}
\accepted{March 27, 2026 }
\submitjournal{ApJ}

\shorttitle{AGN Disks as SN Mufflers I}

\begin{document}

\title{AGN Disks as Supernova Mufflers I: 3D Local Hydrodynamic Models}

\shortauthors{Cook et al.}

\correspondingauthor{Harrison E. Cook}

\author[0000-0001-7163-8712,sname='Cook',gname='Harrison']{Harrison E. Cook}
\affiliation{Department of Astronomy, New Mexico State University,
1780 E University Ave,
Las Cruces, NM 88003, USA}
\email[show]{hecook@nmsu.edu}

\author[0000-0002-3768-7542]{Wladimir Lyra}
\affiliation{Department of Astronomy, New Mexico State University,
1780 E University Ave,
Las Cruces, NM 88003, USA}
\email{wlyra@nmsu.edu}

\author[0000-0003-0064-4060]{Mordecai-Mark Mac Low}
\affiliation{Department of Astrophysics, American Museum of Natural History,
200 Central Park West, New York, NY 10024}
\email{mordecai@amnh.org}

\author[0000-0002-5956-851X]{K. E. Saavik Ford}
\affiliation{Department of Astrophysics, American Museum of Natural History,
200 Central Park West, New York, NY 10024}
\affiliation{Department of Science, BMCC, City University of New York, New York, NY 10007}
\affiliation{Graduate Center \& CUNY Institute for Astronomy (CIfA), City University of New York, 365 5th Avenue, New York, NY 10016, USA}
\email{sford@amnh.org}

\author[0000-0002-9726-0508]{Barry McKernan}
\affiliation{Department of Astrophysics, American Museum of Natural History,
200 Central Park West, New York, NY 10024}
\affiliation{Department of Science, BMCC, City University of New York, New York, NY 10007}
\affiliation{Graduate Center \& CUNY Institute for Astronomy (CIfA), City University of New York, 365 5th Avenue, New York, NY 10016, USA}
\email{bmckernan@amnh.org}

\begin{abstract}

Supernova (SN) shocks that originate from stars on orbits embedded in dense active galactic nuclei (AGN) accretion disks evolve differently from those that occur in the interstellar medium. 
We aim to assess how shocks evolve in this dense stratified medium and understand where SNe are muffled and have their kinetic energy absorbed by an AGN disk versus escaping. We use Sirko \& Goodman (SG) and Thompson, Quataert \& Murray (TQM) AGN disk models for midplane radial profiles, generated with the pAGN code; we compare the disk pressure to the energy of a standard core-collapse SN ($10^{51}\,{\rm erg}$) to find radii where shock breakout can occur.
For verification, we evolve three-dimensional hydrodynamic shearing box simulations of stratified Gaussian disks constructed from the midplane values that are injected with energy and mass from SNe placed at multiple radii and vertical locations, using the Athena code.
We find SN shocks in SG disks around black holes with mass $\Mbh=10^6\,\Msun$ become muffled beyond $R\sim10^6\,\Rs$, and that this muffling radius is inversely proportional to supermassive black hole (SMBH) mass with muffling occurring at $R\sim10^2\,\Rs$ for $\Mbh=10^9\,\Msun$.
Around TQM disks, the muffling radius occurs at $R\sim10^6\,\Rs$, independent of $\Mbh$.
The largest determining factor for muffling a SN shock is the local scale height of the AGN disk.
In conclusion, we developed a predictive analytic criterion to identify where AGN disks can muffle SNe shocks depending on their density and vertical scale.

\end{abstract}


\section{Introduction} 
\label{sec:intro}

Galactic nuclei are typically composed of a SMBH and a nuclear star cluster \citep{Ghez+98,Ghez+08}.
When gas accretes onto the SMBH, some of the objects in the now active galactic nucleus will be embedded in the accretion disk while others will have disk-crossing orbits.
A hypothesis has been suggested that the disk may generate chains of hierarchical mergers between stellar-mass black holes that produce intermediate-mass remnants \citep{McKernan+14, Bartos+17, Stone+17, FordMcKernan25, Abbott+2020-GW190521, LVK+25-GW231123}, much like the collisional growth model for planet formation in circum-stellar disks \citep{Pollack+96}.
These mergers could produce measurable electromagnetic counterparts from interactions with the AGN gas which would be otherwise absent for two black holes merging in the field \citep{McKernan+19}.
Yet, other energetic events such as core-collapse SNe could become a source of noise that must be accounted for in the search for these merger counterparts \citep{Graham+20}.

In addition to those stars initially embedded in the disk when it forms, self-gravitating AGN disks can naturally fragment and form new stars, particularly at large radii \citep{Shlosman89, Collin08}, though magnetic fields \citep{Gerling-Dunsmore+25, Tsung+25} and feedback from embedded objects \citep{HanklaJiangArmitage20} may inhibit collapse.
Furthermore, AGN disks can quickly capture stars from the inclined population \citep{Fabj20,MacLeod20}.
Any massive embedded stars may detonate as SNe within the disk confines, though stellar evolution in AGN disks may proceed very differently than in the interstellar medium \citep[ISM,][]{Cantiello21}.
Even in this case, SNe may yet result from stellar evolution \citep{Jermyn21} or collisions.
SNe in the dusty outskirts of AGN disks may have already been detected as bright optical and infrared flares \citep[e.g.][]{Assef18}.

We are motivated to explore the evolution of SNe embedded in an AGN disk due to the possibility that AGN flares could be electromagnetic counterparts to the gravitational wave detections from binary black hole (BBH) merger events, one particular case to underscore being GW190521 \citet{Abbott+2020-GW190521}.
While the observed flare \citep{Graham+20} did not resemble a standard SN light curve, embedded SNe could behave differently than SNe occurring in the ISM. 
Thus, by characterizing SN light curves in AGN, we can rule out false positive counterparts to BBH mergers as well as constrain the rate of SNe in AGN. 
Looking forward, detections of embedded SNe (or other transients in AGN disks) will allow us to constrain the composition, evolution, and dynamics of nuclear star clusters that coexist with an AGN disk, as well as to probe the disk properties.

Whether the radiation from SNe escapes from an AGN disk is a separate matter from the mechanical breakout of the shockwave.
The non-uniform density and variable disk thickness have large impacts on a SN shock's morphological in comparison to their galactic counterparts \citep{MacLowMcCray88}.
The right combination may slow down a SN shock significantly or prevent it from escaping entirely, thus depositing their kinetic energy in the disk.
These muffled SNe may not punch holes through the disk but could still be a contributing energy source for driving turbulence and maintaining pressure support in disks that would otherwise collapse under self-gravity \citep{Rozyczka+95,Moranchel+21}.

\citet{Grishin+21} performed two dimensional hydrodynamic simulations in the $r$-$z$ plane of SNe occurring in a flat slab of gas at the midplane and with a vertical offset and find the shock breaks free from the disk above and below the midplane for both cases.
\citet{Moranchel+21} ran three-dimensional simulations in spherical coordinates for the upper half of the disk and find agreement that shocks from SNe at the midplane broke free from the disk.
We expand upon this work by estimating where SNe may break free from two AGN disk models and perform three-dimensional hydrodynamic simulations of a SN embedded in a vertically stratified shearing disk for midplane and offset SNe explosions.
In this work we describe the hydrodynamic model and will present the radiative transfer analysis in a companion paper.

This paper is organized as follows.
Section 2 details predictions for SN breakout and our hydrodynamic models with results described in Section 3.
We discuss he implications of our findings in Section 4 and highlight final conclusions in section 5.


\section{Methods} \label{sec:methods}

\citet{MacLowMcCray88} showed repeated SNe from star clusters can inflate superbubbles in disk galaxies.
These bubbles can break free from the disk when the internal mechanical wind luminosity exceeds a characteristic luminosity scale set by local intensive thermodynamic quantities.
Using their formalism, we can analytically compute where SNe are expected (or not expected) to break out from the AGN disk with further implications for observability dependent on the opacity.

Whether a SN explosion occurring at an AGN disk's midplane will break free from the disk or be muffled, can be roughly determined using the ram pressure of the SN, or, equivalently, its energy density
\begin{equation}
    \varepsilon_{\rm SN} \equiv \frac{\Esn}{V_{\rm SN}}\ {\rm erg\ cm^{-3}}
\end{equation}
and the gas pressure of the disk at the explosion site
\begin{equation}
    P_D(r) = n k T.
\end{equation}
The number density $n$ and temperature $T$ are those of the disk at the explosion's radial distance from the SMBH.
Here, $k$ is Boltzmann's constant, $\Esn$ is the total kinetic energy released by the SN (typically, $10^{51}$ erg), and $V_{\rm SN}$ is the volume bound by the SN shock wave.
The relative influence of the two is determined by their ratio
\begin{equation}
    \C \equiv \frac{\varepsilon_{\rm SN}}{P_{\rm D}}.
\end{equation}

At early stages of the SN evolution, the shock will remain quasi-spherical, and its volume after expanding to a radius $\Rsn$ is thus $V_{\rm SN} = (4\pi/3) \Rsn^3$ leading to
\begin{equation}
    \C = \frac{\Esn}{V_{\rm SN}P_{\rm D}} = \frac{3}{4\pi k} \frac{\Esn}{n T R_{\rm SN}^3}.
\end{equation}
We set the SN radius at breakout to be one disk scale height $\Rsn=H$, assume a pure hydrogen gas (proton mass $\mProt$), and substitute the gas volume density $\rho=n\mProt$. The criterion for breakout becomes 
\begin{equation} \label{eqn:C}
    \C = \frac{3\mProt}{4\pi k}\frac{\Esn}{\rho \,T\,H^3}.
\end{equation}
Though the vertical structure is Gaussian, we set $\rho$ and $T$ to the midplane quantities as conservative estimates.

\begin{figure}
    \centering
    \includegraphics[width=1.0\linewidth]{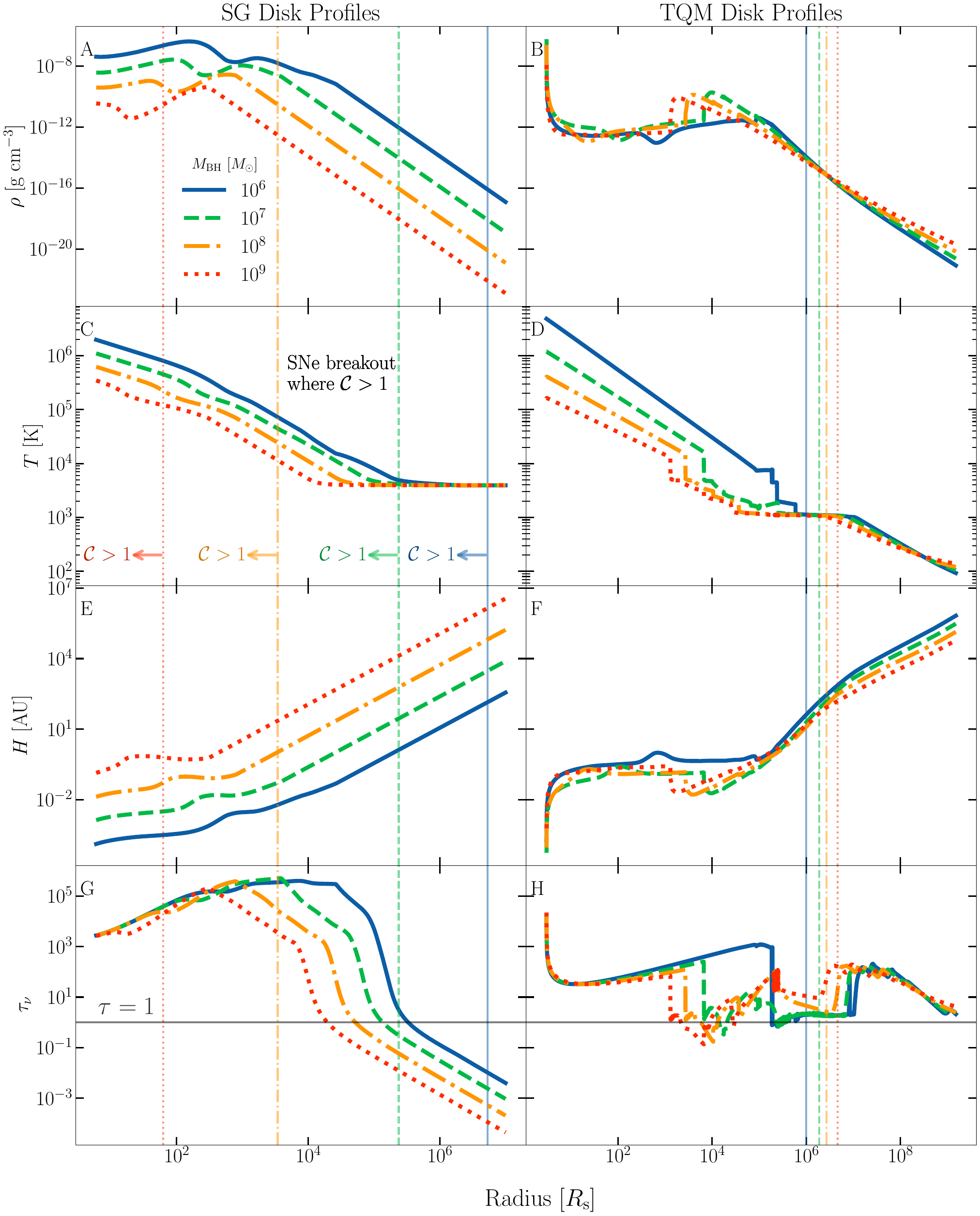}
    \caption{\citet{SirkoGoodman03} (left column) and \citet{Thompson+05} (right column) accretion disk models of midplane density (A, B), midplane temperature (C, D), local scale height (E, F), and optical depth (G, H) as functions of radius for disks orbiting various SMBH masses.
    Vertical lines indicate where $\mathcal{C}=1$ for accretion disk models corresponding to SMBH with the same line style.}
    \label{fig:disk-models}
\end{figure}

When $\C > 1$, the pressure in the SN shock exceeds the disk's gas pressure, and the SN will break free from the disk.
Since the path of least resistance lies perpendicular to the disk midplane, most of the expansion will be directed along the $z$-direction, temporarily excavating a column of the disk until the disk shear elongates and fills the cavity \citep{Rozyczka+95, Moranchel+21}. 
When $\C < 1$, the SN lacks the pressure to break free from the disk, leading to a \emph{muffled} SN.
Because they do not completely perforate the disk, these SNe impart a significant portion of $\Esn$ as heat into the disk, modifying the local temperature and disk structure.
In the left column of Fig.~\ref{fig:disk-models} we show radial profiles of midplane density, midplane temperature, scale height, and optical depth for SG disks orbiting SMBHs with masses $M_{\rm BH}=\{10^6, 10^7, 10^8, 10^9\}\,\msun$ as calculated with the \texttt{pAGN} package \citep{Gangardt+24}.

For a SN explosion with $\Esn=10^{51}$ erg orbiting a $10^8\,\msun$ SMBH at 1000 Schwarzschild radii $\Rs$ in an SG disk, the midplane temperature, density, and local scale height are $T\sim 10^5$ K, $\rho=10^{-9}\ \gcmcube$, and $H \sim 10$ AU, respectively.
We find $\C \sim 2600$, indicating a strong breakout is expected.
Substituting these values we can rewrite $\mathcal{C}$ as a scaling relation:

\begin{eqnarray}
    \C &\approx& 2600 \ \left(\frac{\Esn}{10^{51}\,{\rm erg}} \right) \ \left({\frac{\rho_0}{10^{-9}\,\rm{g\,cm^{-3}}}}\right)^{-1}  \nonumber \\
    &&\left({\frac{T}{10^5\,{\rm K}}}\right)^{-1} \
    \left({\frac{H}{10\,{\rm AU}}}\right)^{-3} \ 
    \exp\left( \frac{z^2}{2H^2}\right)
    \label{eqn:C-parameterized}
\end{eqnarray}
where we have also introduced the midplane density $\rho_0$ and the exponential vertical profile in terms of $H$ to accommodate SNe on inclined orbits (see Sec.~\ref{sec:disk-model} for more details).

Profiles for the TQM disk model in the right column of Fig.~\ref{fig:disk-models} tell a different story resulting in a drastically altered landscape for $\C$ as discussed later and shown as a function of BH mass and radius in Fig.~\ref{fig:breakout-midplane}.
For all SMBH masses, the inner TQM disk is less dense while the opposite is true in the outer torus.
The temperatures are similar to the SG model in the inner disk, but where SG bottoms out at $\sim 4 \times 10^3\,\rm{K}$ and holds that temperature until the disk edge, the TQM model extends down to $100\,\rm{K}$.
The density differences are reflected in the local scale height, which are elevated in the inner regions, while suppressed beyond $R>10^5\,\Rs$.
The optical depths behave much differently, with SG disks typically transitioning from being optically thick to thin at $R=10^5\,\Rs$ and TQM remaining entirely optically thick except near $R=10^4\,\Rs$ for $\Mbh<10^7\,\msun$.
The vertical lines indicate where $\mathcal{C}=1$ for disk models corresponding to the same line styles indicating the SMBH mass.

\subsection{Hydrodynamic Models}
\label{sec:hydro}

We test the validity of Eq.\,(\ref{eqn:C-parameterized}) using three-dimensional shearing box hydrodynamic simulations with the publicly available Athena code, a finite-volume code for compressible hydrodynamics problems \citep{Stone+08, StoneGardiner10}.
The code solves the Riemann problem at each cell interface to determine the flux of quantities over time.
We find the HLLE solver \citep{Einfeldt+91} achieved the most robust numerical stability across the strong shocks we produce and is employed for all our models.

\subsection{Dynamical Equations} \label{sec:dyn-eq}
Athena solves the equations of motion in the standard Cartesian shearing box approximation in the following form
\begin{align}
    &\ptderiv{\rho} + \Div{\left(\rho\vec{}{v}\right)} = 0
    \label{eqn:cons_mass} \\
    &\ptderiv{\rho \vec{v}} + \Div{\left(\rho \vec{v} : \vec{v}\right)} + \nabla P =
    \rho \Omega_{0}^{2}(2qx{\bf \hat{i}} - z{\bf \hat{k}}) - 2\Omega_{0} {\bf \hat{k}} \times \rho \vec{v}
    \label{eqn:cons_momentum} \\
    &\ptderiv{E} + \Div{\left[\vec{v}\left(E + P\right)\right]} =
    \Omega_{0}^{2}\rho\vec{v} \cdot (2qx{\bf \hat{i}} - z{\bf \hat{k}}).
    \label{eqn:cons_energy}
\end{align}
where $\rho$ is the gas density, $t$ is time, $\vec{v}$ the gas velocity, $P$ is the gas pressure, $\Omega(r) = (GM_{\rm BH}/r^3)^{1/2}$ is the Keplerian orbital frequency evaluated at the radial location corresponding to the center of the shearing box.
The logarithmic shearing parameter $q$ is related to the orbital frequency by
\begin{equation}
    q \equiv -\frac{d\ln\Omega(r)}{d\ln r}.
\end{equation}
We set $q=3/2$ for a Keplerian flow.
The variables $x$, $y$, and $z$ represent the local radial, azimuthal, and vertical coordinates within the domain, respectively.
The total energy density is the sum of the internal and kinetic energy densities
\begin{equation}
    E = \frac{P}{1-\gamma} + \frac{1}{2}\rho v^2,
\end{equation}
where $v \equiv \sqrt{ \vec{v} \cdot \vec{v} }$ is the velocity.

\subsection{Hydrodynamic Initial Conditions} \label{sec:ICs}

\subsubsection{AGN Disk Model} \label{sec:disk-model}

We assume an initial state of hydrostatic equilibrium resulting in a vertical disk density profile with the form
\begin{equation} \label{eqn:disk-z-profile}
    \rho_{\rm D} (z) = \rho_0 \, \exp\left[-\frac{\gamma}{2} \left(\frac{\Omega_0 z}{\cs}\right)^2 \right]
\end{equation}
where $\rho_0$ is the disk midplane density at our domain's radial location in the SG disk. We use the disk aspect ratio $h=H/r$ at the selected radius to find the scale height $H$ and thus the sound speed $\cs=\Omega H$, which we set to be constant throughout the domain in the initial conditions.
For the purposes of this paper, we only adopt values from the SG disk model for our hydrodynamic models.

\subsubsection{Initializing the Supernova}
We model the SN as a symmetric three-dimensional Gaussian perturbation in density and energy centered on coordinates $(x_0, y_0, z_0)$.
In our models, $(x, y, z) = (0, 0, 0)$ is the center of the shearing box domain, and $z=0$ corresponds to the disk midplane.
We set $z_0=H$ when exploring SNe detonating away from the disk midplane.
We distribute the additional density perturbation from the ejecta mass $\rhoej$ into each zone with
\begin{eqnarray}
    &\Delta r^2 = (x-x_0)^2 + (y-y_0)^2 + (z-z_0)^2, \label{eqn:sn-radius}\\
    &\rhoej (x, y, z) = \rho_{\rm a}\,  {\rm exp}\left(-\frac{\Delta r^2}{2 \Rsn^2}\right). \label{eqn:ejecta-gauss}
\end{eqnarray}
The density amplitude is set by
\begin{equation} \label{eqn:ejecta-amplitude}
    \rho_{\rm a} = \frac{\Mej}{(2 \pi)^{3/2} \Rsn^3}. 
\end{equation}
The total ejecta mass $\Mej$, which we set to $10\,\msun$, is normalized by integrating \eq{eqn:ejecta-gauss} with \eq{eqn:sn-radius} and \eq{eqn:ejecta-amplitude} over all space.

We distribute the SN perturbation in energy density in the same manner with
\begin{equation} \label{eqn:energy-gauss}
    \esn (x, y, z) = \varepsilon_{\rm a}\, \exp\left(-\frac{\Delta r^2}{2 \Rsn^2}\right)
\end{equation}
and amplitude
\begin{equation} \label{eqn:energy-amplitude}
    \varepsilon_{\rm a} = \frac{ \Esn }{ (2 \pi)^{3/2} \Rsn^3 }. 
\end{equation}
Here $\Esn = 10^{51}$ erg is the total initial kinetic energy budget of a core-collapse SN.
During testing before adding the disk, we find source regions smaller than 6 zones cause the grid to imprint a cubic shape onto the shock rather than allowing it to expand spherically symmetrically
Thus, we set the initial SN's radius $\Rsn = 3\Delta x$ in \eq{eqn:ejecta-gauss} where $\Delta x$ is the local cell size.

\citet{Simpson+15} showed one must consider the time $\tpds$ and radius $\Rpds$ at which the transition from the Sedov-Taylor phase to the pressure-driven snowplow (PDS) phase occurs when adding the SN's energy to the domain. This transition depends on a combination of the density of the ambient gas and the local cell size and informs the fraction $\fkinetic$ of the total initial kinetic energy budget of the SN that remains in the kinetic phase $\varepsilon_{\rm K} = \fkinetic\,\esn$ and the fraction that has been converted to internal energy $\varepsilon_{\rm I} = (1-\fkinetic)\,\esn$. If $\Rpds < \Delta x$, the transition is unresolved and the SN can be added entirely as internal energy (i.e., $\fkinetic = 0$). But if $\Delta x < \Rpds$, the transition is resolved and some fraction of the energy must be added as kinetic energy (i.e., $0\leq\fkinetic\leq1$), which we calculate according to Section 3.5 in \citet{Simpson+15}. In practice, we find an upper limit of $\fkinetic \leq 0.16$ to avoid the so-called carbuncle instability \citep{PeeryImlay88, Robinet+00} that leads to numerical problems that eventually terminate the simulation.

\subsection{Computational Domain} \label{sec:HydroDomains}

The shearing box domain is characterized by shear-periodic boundaries in the $x$ (radial) and $y$ (azimuthal) dimensions.
In addition, we use outflow boundary conditions for the $z$ (vertical) direction.
The azimuthal flows leave and re-enter the periodic boundaries representing the flow of a continuous disk.
Any shocks that enter the domain after passing through these periodic boundaries would have been emitted by adjacent SNe and produce effects in our domain no longer be consistent with an isolated SN.
As such, we stop the calculations when any material ejected from the disk reaches these boundaries, and the shocks expanding through the midplane never approach the periodic boundaries during our models.
For calculations performed at $R=10^3\,\Rs$, our domains span 96x96x128$\,H$ with a resolution of 960x960x1280 zones in the $x$, $y$, and $z$ dimensions, respectively.
At $R=10^5\,\Rs$, the SN does not significantly expand in $z$, so we use a cubic domain that spans $24^3\,H$ at a resolution of 960x960x960.
As shown later, the vertical disk profile shown in Eq.\,(\ref{eqn:disk-z-profile}) causes the shock to expand much more rapidly in the vertical direction than horizontally.
Thus, our domains are constructed such that breakout from the disk occurs well before horizontal flows through the midplane could cross the $x$ or $y$ boundaries and break the assumption of an isolated SN for the run.


\section{Results} \label{sec:results}

\subsection{Breakout Parameter}
\begin{figure}
    \begin{subfigure}{0.5\textwidth}
    \centering
        \includegraphics[width=1\linewidth]{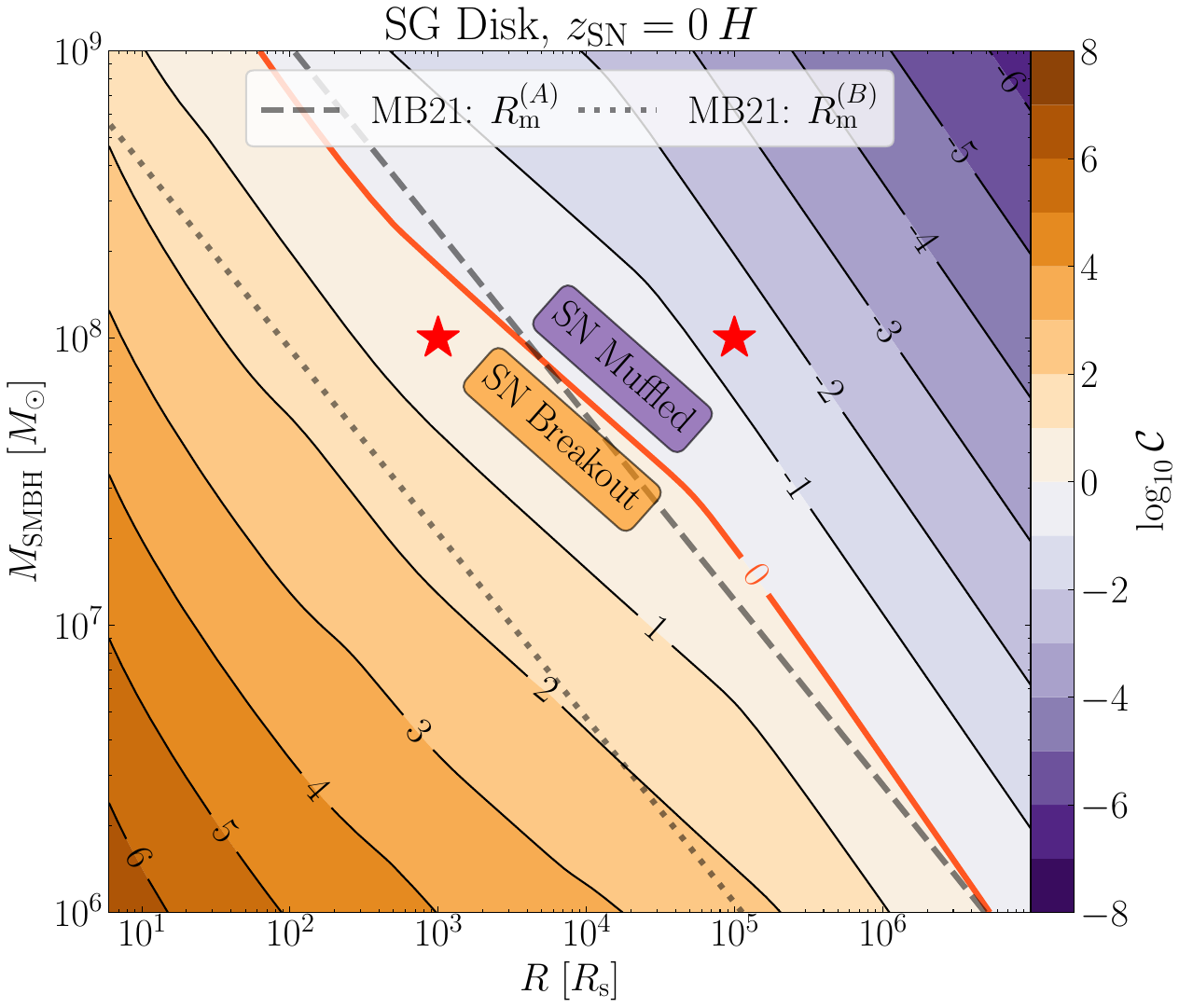}
    \end{subfigure}
    \begin{subfigure}{0.5\textwidth}
    \centering
        \includegraphics[width=1\linewidth]{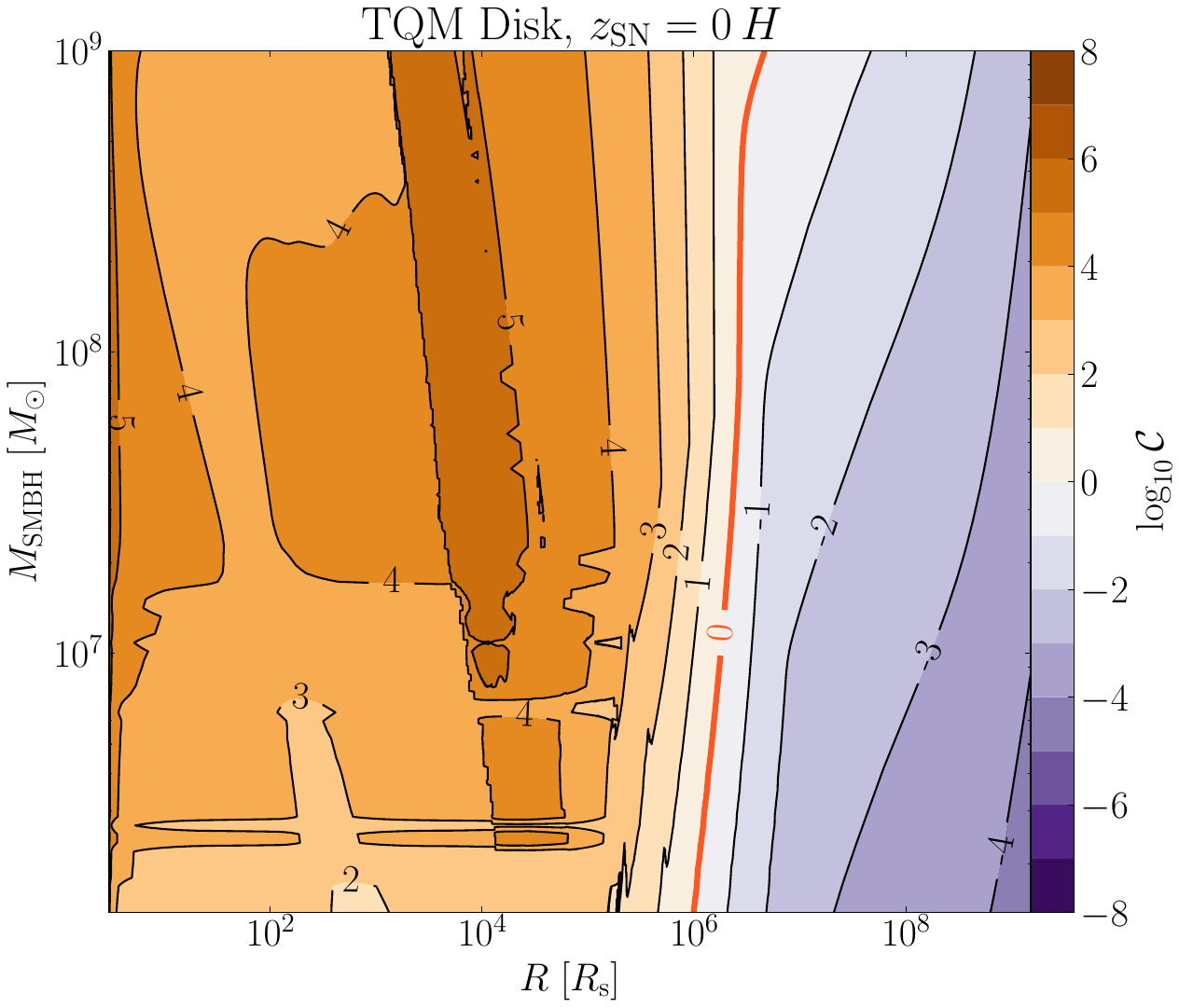}
    \end{subfigure}
    \caption{Breakout parameter $C$ versus SMBH mass and radius in the accretion disk for SNe detonating at the disk midplane.
    \emph{Top Panel:} In SG disks around SMBH with mass $10^6\,\msun$, SNe with energy $10^{51}\,{\rm erg\ s}^{-1}$ will strongly break free from all radii $\leq5\times10^6\,\Rs$.
    The muffling radius decreases in $\Rs$ as SMBH mass increases, reaching $10^2\Rs$ for those with $M=10^9\,\Msun$.
    Stars indicate the $\Mbh$ and radial locations we use to set up our hydrodynamic shearing box simulations to test this prediction.
    The dashed line marks where \citet{Moranchel+21} calculated the muffling radius for SNe when radiative cooling is not important (their scenario A) which closely matches the line for $\C=1$.
    The dotted line marks their scenario B in which radiative cooling becomes important by the free expansion phase, reducing the pressure interior to the shell.
    \emph{Bottom Panel:} The same for TQM disks. The location of the muffling radius is located at $R_{\rm m}=10^6\,\Rs$ for all SMBH masses.}
    \label{fig:breakout-midplane}
\end{figure}

Fig.~\ref{fig:breakout-midplane} shows the range of values for $\C$ using SG and TQM disk models generated with \texttt{pAGN} for 100 SMBH mass bins logarithmically spaced over the range $10^6 \leq \Mbh \leq 10^9\,\Msun$.
We start the TQM models at $\Mbh=10^{6.14}\,\Msun$ due to mass accretion rate limitations in this model for low-mass SMBHs.
The orange region indicates where $\log_{10}\C>0$, meaning SNe with $10^{51}$ erg of energy will break free from the disk.
The purple region indicates where $\log_{10}\C<0$ and these SNe will be muffled.
We refer to the boundary between these two regions as the muffling radius $\Rb$.
The stars in the figure indicate the locations of our hydrodynamic models in this phase space whose results we describe in the next section.
Overplotted are muffling radii as calculated by \citet{Moranchel+21} for scenarios when radiative cooling decreases the SN remnant's internal pressure ($\Rb^{(B)}$) and when it is not ($\Rb^{(A)}$).

Around an SMBH with $M_{\rm BH}=10^6\,\msun$, the shock will break free from the disk at all radii out to $5\times10^6\,\Rs$.
Shocks from SNe beyond this radius out to the disk edge at $10^7\,\Rs$ will be muffled and not break free.
The radius of this muffling boundary is inversely proportional to SMBH mass, decreasing to $\sim 6\times10^2\,\rs$ for $\Mbh=10^9\,\msun$.
For SG disks around SMBHs with mass $10^9\,\Msun$, $H$ is consistently $10^3$ to $10^4$ times larger at all radii than for disks around those with masses of $10^6\,\Msun$ (c.f., Fig.~\ref{fig:disk-models}).
Eq.\,(\ref{eqn:C}) most strongly depends on the scale height ($\propto H^{-3}$).
Thus, puffier disks are more likely to obscure embedded SNe than thin disks.

Beyond $R=10^3\,\Rs$ for disks around $\Mbh=10^8\,\Msun$, the SG model is unstable to collapse since Toomre's stability parameter $Q\approx\Omega^2/2\pi G\rho < 1$.
To counteract this, they artificially set $Q=1$ by invoking some unknown source of energy that is possibly tied to in-disk star formation that makes the disk beyond this radius critically stable against gravitational collapse.
For most SMBH masses the entire disk will allow SNe with $\mathbb{E}_{\rm SN}\geq10^{51}$\,erg to break free.
At $\Mbh> 2\times10^8\,\Msun$, the muffling radius shrinks down to $R\approx100\,\Rs$ when $\Mbh=10^9\,\Msun$.
Since much of the disks will muffle SNe, they could be an additional component for maintaining pressure support.

The results for the TQM disk (bottom panel of Fig.~\ref{fig:breakout-midplane}) demonstrate a different behavior.
The muffling radius has no dependence on the SMBH mass such that $\C(10^6\,\Rs)=1$ for all masses.
The range of values for $\C$ is more confined than the SG disk and ranging over $10^{-4}\leq \C \leq 10^5$.
Additionally, the TQM disk model produces a peak in the breakout zone around $R=10^4\,\rs$ where $\C$ briefly reaches a local maximum before dropping off again at larger radii.
Furthermore, for $\Mbh>10^7\,\msun$, this peak reaches $\log_{10}\C>5$, which decreases for lower SMBH masses.
Finally, TQM disks around $\Mbh\sim10^6$ have relatively constant $\C$ at $10^2$ up to about $R\sim10^5\,\Rs$ before dropping towards the muffling radius at $10^6\,\Rs$, congruent with the general behavior.
This feature corresponds to a trough in the scale height in the TQM disk model due to sudden drops in temperature at this radius (right column of Fig.~\ref{fig:disk-models}), a feature missing from the SG disk model.

\begin{figure}
    \centering
    \includegraphics[width=1\linewidth]{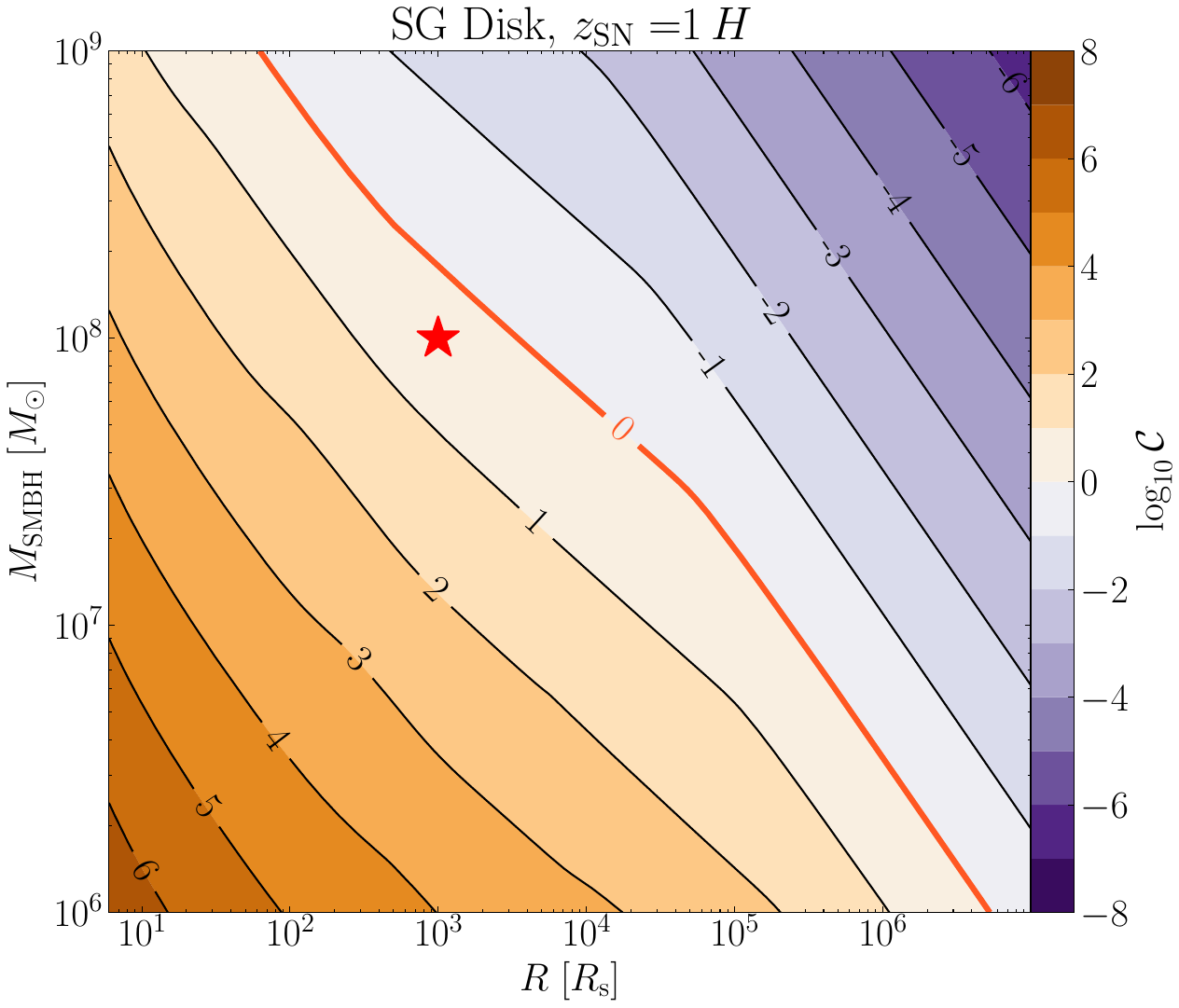}
    \caption{Breakout parameter $\C$ for SNe at height $z=1\,H$ for an SG disk.
    The lower density offers less resistance to the shock expansion, making breakout easier, and pushing the muffling boundary outward compared to midplane detonations in Fig.~\ref{fig:breakout-midplane}.
    The star marks the location that our off-midplane run samples this analytic prediction.}
    \label{fig:breakout-sg-z1h}
\end{figure}

Stars can be on inclined orbits meaning they need not lie precisely at the midplane when a SN occurs.
Fig.~\ref{fig:breakout-sg-z1h} shows the breakout value $\C$ for SNe detonating in SG disks at a height $z=1\,H$.
The muffling radius, where $\log_{10}\C=0$, shifts outwards in comparison to midplane SNe.
The reduced density away from the midplane offers less resistance, so shock breakout away from the midplane occurs more easily.

\subsection{Hydrodynamics}

\begin{figure*}
    \centering
    \includegraphics[width=1.0\textwidth]{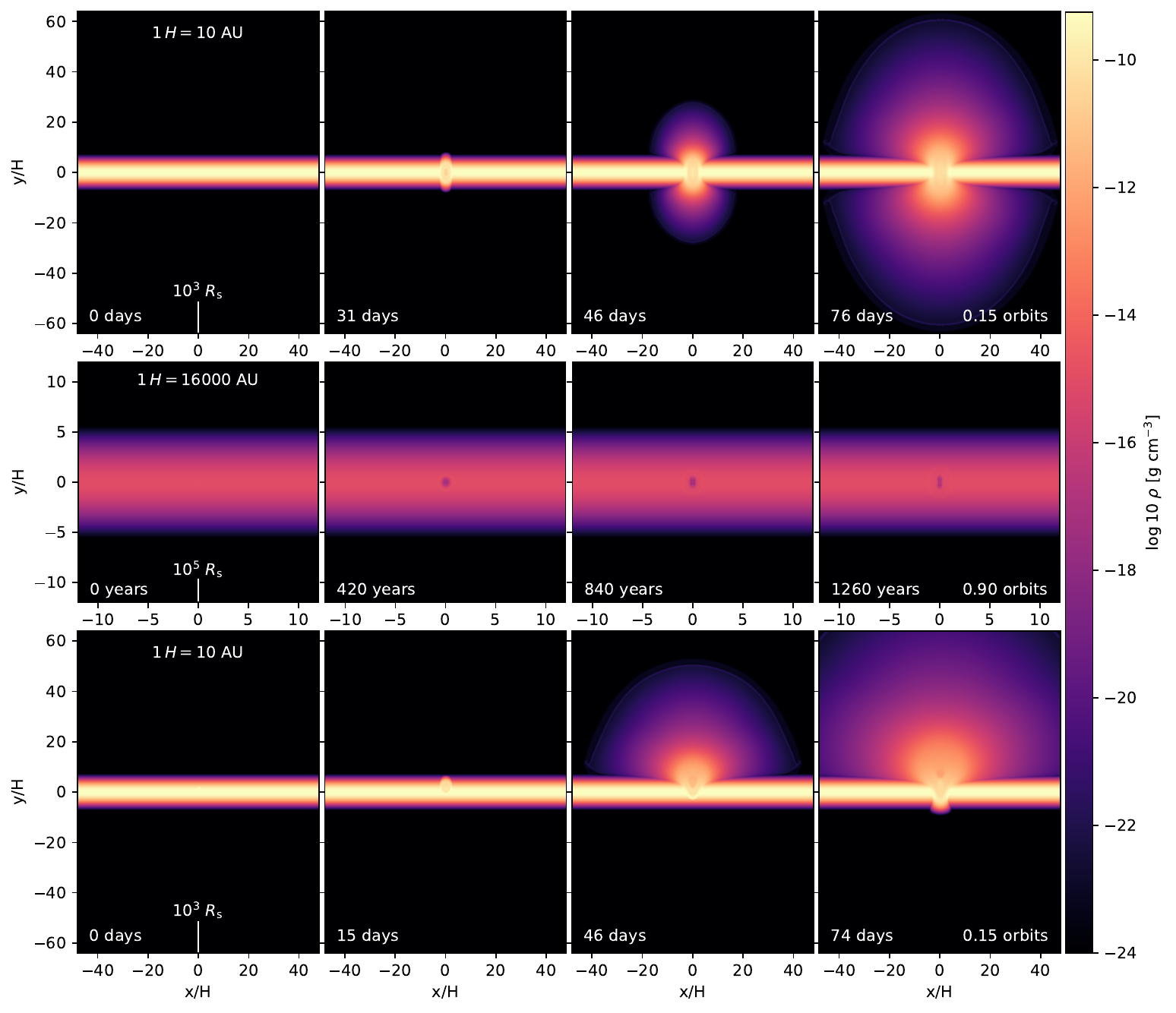}
    \caption{Time series of vertical slices ($x-z$) showing the gas volume density evolution for SNe embedded in accretion disks around an SMBH with mass $M=10^8\,\msun$.
    \emph{Top row}: Midplane detonation at $R=10^3\,\Rs$.
    \emph{Middle row}: Midplane detonation at $R=10^5\,\Rs$.
    \emph{Bottom row}: Offset detonation ($z=1H$) at $R=10^3\,\Rs$.}
    \label{fig:timeseries-composite-xz}
\end{figure*}

We test the muffling radius $Rb$ marked in red in Fig.~\ref{fig:breakout-midplane} for a disk orbiting a $10^8\,\msun$ SMBH by placing our hydrodynamic shearing boxes at $R=[10^3, 10^5]\,\rs$, which have breakout parameters $\C=[2.6, 0.04]$, respectively.
These parameters correspond to the stars in that figure.
The top row of Fig.~\ref{fig:timeseries-composite-xz} shows a time series for a SNe at $R=10^3\,\Rs$ and the disk midplane $z_0=0$.
In early times, the shock descends the pressure gradient of the disk and primarily expands vertically until it reaches $z=5\,H$ after 30 days where it is free to expand horizontally along the disk surface.
At this point, the hemispheric expansion evolves symmetrically about the midplane and reaches our domain boundaries at $z=\pm65\,H$ in $\sim80$ days.
The middle row of Fig.~\ref{fig:timeseries-composite-xz} shows the SN placed at $R=10^5\,\rs$ and at the disk midplane.
In contrast to the SN at $10^3\,\rs$, this one remains muffled, only reaching $z\sim1\,H$ over the 1200 years evolution.
While the SN shock is physically contained at $R=10^5\,\Rs$, the SG disk model is optically thin at this location for $\Mbh \geq 10^7\,\msun$ leaving the possibility that the emitted photons still escape.
The disks' radial range in each setup span the same number of scale heights $H$ but the full domains encompass drastically different volumes since $H=10$ AU vs. 16,000 AU, respectively.
Stellar winds from massive SNe progenitors can carve cavities in the low density gas in the outer disk \citep{Fu-Lin+23}, but the large scale height at $R=10^4\,\Rs$ leaves them unresolved in our models.

The SNe in these two runs occurred directly at the disk midplanes, but in practice the dynamical nature of the AGN system means a star's orbit can be inclined relative to the accretion disk.
The bottom row of Fig.~\ref{fig:timeseries-composite-xz} shows a time series for a setup identical to the top row except we offset the SN initialization vertically to $z_0=1\,H$.
With such a large head start, the upward propagating shock reaches the disk surface in half the time as the midplane case; likewise for the time to reach the boundary.

The plume is slightly more spherical than the midplane case, which directs more of the energy perpendicular to the disk.
The downward shock eventually crosses the midplane and exits the far side of the disk, but the additional material between the SN and midplane slows down the flow which becomes subsonic by the time it reaches the midplane.
We have extended the time series for this run beyond shock-boundary interaction in order to show the flow traveling through the disk.
By allowing the flow to exit the horizontal periodic boundaries, the assumption of a single SN in the disk breaks, but the shock interactions remain at those boundaries and do not affect the source region.

\begin{figure*}
    \centering
    \includegraphics[width=1.0\textwidth]{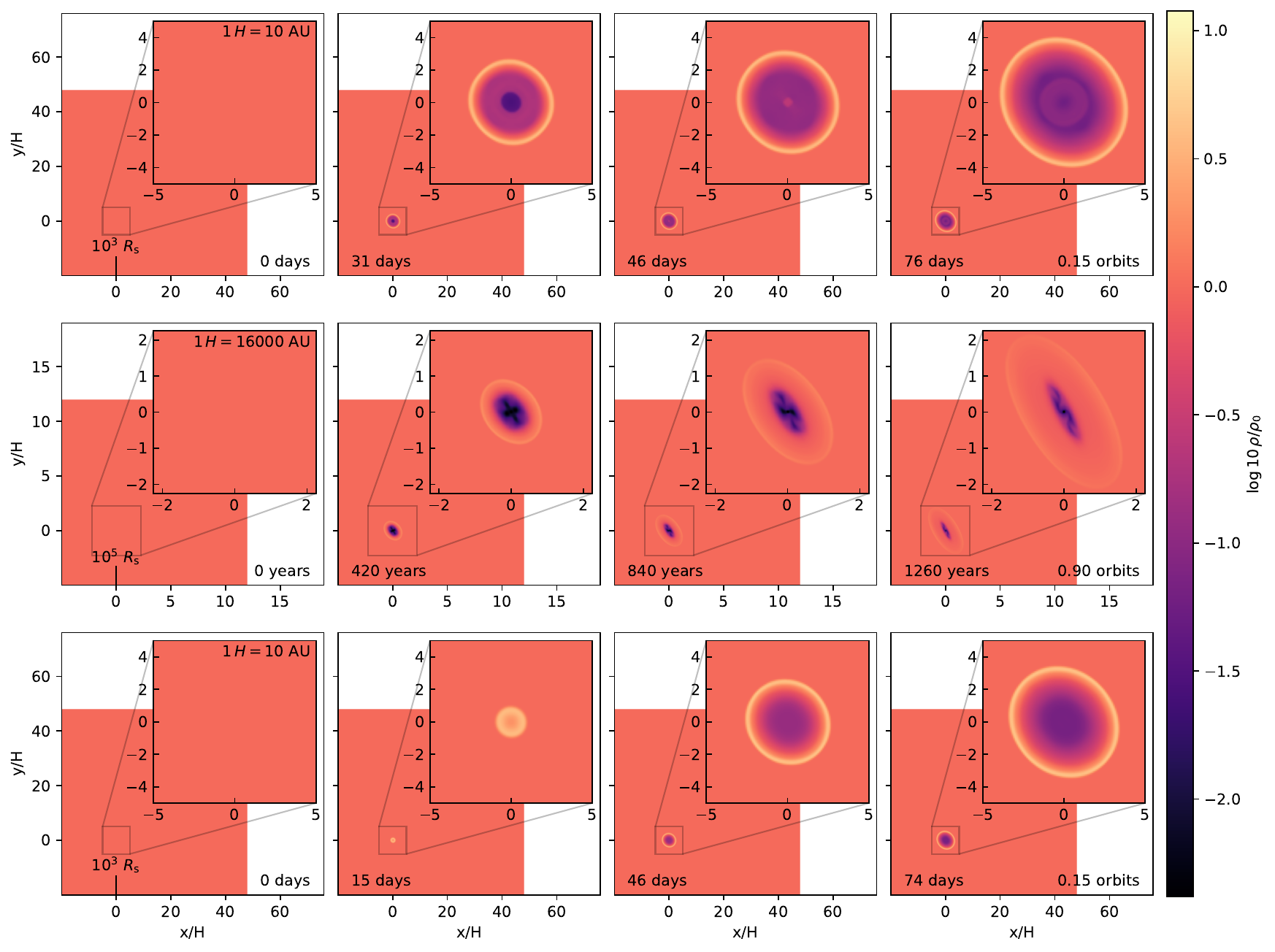}
    \caption{Time series of midplane slices ($x-y$) showing the gas volume density evolution for SNe embedded in accretion disks around an SMBH with mass $M=10^8\,\msun$.
    \emph{Top row}: Midplane detonation at $R=10^3\,\Rs$.
    \emph{Middle row}: Midplane detonation at $R=10^5\,\Rs$.
    \emph{Bottom row}: Offset detonation ($z=1H$) at $R=10^3\,\Rs$.}
    \label{fig:timeseries-composite-xy}
\end{figure*}

Fig.~\ref{fig:timeseries-composite-xy} shows the midplane density fields normalized by their respective initial densities.
The main figures in each panel are plotted using the same snapshots and in their native scales (960x960 zones) matching Fig.~\ref{fig:timeseries-composite-xz}, but they have been shifted down and to the left here to accommodate the insets which zoom in on the SNe.
For the midplane run at $R=10^3\,\Rs$ (top row), the shock expands symmetrically until the disk's shear begins to take affect by 0.15 orbits, causing it to elongate slightly along the $y$-direction.
The muffled midplane run at $R=10^5\,\Rs$ (middle row) runs for 0.9 orbits.
Over this longer timescale, the shear effects become more pronounced.
The shock moving leftward towards the SMBH encounters annuli of gas moving at faster orbital velocities which sweep up the shock in a tail wind and pull it forward.
Similarly, the rightward moving shock encounters slower moving gas and conversely lags behind.
Over time, this shearing elongates the cavity made by the SN enough to close it, and shrinks its width almost entirely after 420 years.
The final row shows the SN at $z=1\,H$ reaches the midplane within 15 days, which shears accordingly over the next 60 days.
Recall, the downward flow is subsonic by this point, indicating material that pierces the midplane from an offset SN is not guaranteed to break through as a shock.

\section{Discussion} \label{sec:discussion}

Whether a SN that occurs at the midplane of an AGN disk breaks free or is muffled depends on its radial location, the mass of the SMBH, and the disk model -- chiefly the local density, temperature, and scale height.

When a SN occurs away from the midplane, the ambient density is lower and the distance the shock must travel in order to break out is smaller than for a midplane detonation at a given radius with the same energy.
This results in shock breakout occurring sooner and retaining more of its initial kinetic energy.
Conversely, the shock propagating towards the far side of the disk must climb the density and pressure gradient of the disk in order to reach the midplane.
In a relatively low density disk, the flow may remain supersonic as it exits the opposite side, but if the disk is dense enough, or the path length to the other side is long enough, the flow may become subsonic well beforehand.
The exact nature of this asymmetrically propagating shock has strong implications for whether radiation reaches two observers located on opposite sides of the disk.
Similar behavior is seen in \citet{Grishin+21}.

As the disk captures stars \citep{MacLeod20,Nasim+23} some orbits will begin relatively aligned with it while others will start on highly inclined orbits.
By nature of the geometry, the star will always be embedded at the disk at the nodes where the two orbits intersect, and for circular orbits the star will be outside of the disk 90 degrees from the nodes when the inclination is large enough.
The inclination at which the entire orbit becomes embedded in the disk depends on its semi-major axis and the disk's aspect ratio, but as the orbit aligns, the star will progressively spend more time embedded within the disk and closer to the midplane.
Our models show that the upward shock from an offset SN on an inclined orbit evolves more rapidly than a midplane SN in the same disk at the same radius.
This holds for midplane SNe in a disk with a midplane density equal to the local density of an inclined SN.

To verify this, we placed a SN at the midplane of a disk with a midplane density equal to the density at $z=1H$ in the inclined SN model, i.e., $\rho_0=\rho_{\rm D}(z=1H)\approx5\times10^{-10}$ g cm$^{-3}$ using Eq.\,(\ref{eqn:disk-z-profile}).
The morphology of the plumes that exit the disk are practically identical, but the horizontal expansion into the midplane is slightly larger at corresponding times owing to the lower density.
The vertical shock breakout occurs 6 days ahead of a midplane SN in the denser disk and 10 days behind the inclined SN's, and having been slowed down as it climbed out from the midplane, the flow traveling along the disk surface has a slower velocity, reaching the boundary 13 days behind the inclined SN.

Lateral shock expansion into the midplane is small, $\Delta r\approx 10\,H$ at $10^3\,\Rs$.
The relative size of this cavity to its location is $\Delta r/R = 0.05$.
The evolution time is short relative to the orbital period, so the midplane cavities remain close to circular.
Over long times, on the order of the orbital period, the horizontal shock will stall as the shear elongates the cavity until it closes.
Thus, a single SN will have minimal impact on the disk structure and the cavity will close as it stalls and the shear deforms it.

If AGN disks with structures described by the SG model are common, then for a fiducial $\Mbh=10^8\,\Msun$, an observed AGN-SN would have occurred within $10^3\,\Rs$ ($<$1000 AU), corresponding to $\sim5$ light days, inside the broadline region that is a few light weeks to typically a light month in size depending on $\Mbh$ \citep{BlandfordMcKee82}.
This means that the response of the broadlines to the new SN illumination would be roughly symmetric.
Conversely, if the SN happens at large radii in a lower density disk, then the response will be infrared-bright.
Thus, the observational details allow us to use SN to infer its location and probe the disk's properties.

If the SN occurs in the inner disk, the orbital time can be short.
The top row of Fig.~\ref{fig:timeseries-composite-xz} occurs in about 15\% of the orbital period ($\sim3$ years $\sim1000$ days) at $R=10^3\,\Rs$ for $\Mbh=10^8\,\Msun$.
SNe occurring here should lead to redshifted or blueshifted asymmetries in the lines due to the Doppler effect since the orbital speed is of the order $10^4$ km s$^{-1}$.
Similar asymmetries have been observed in AGN broadline profiles that are possibly caused by localized inconsistencies in the disk structure \citep[e.g., spiral arms,][]{Schimoia+17}.
An inner-disk SN may be ruled out as the catalyst for an AGN flare if the spectrum does not exhibit these features.

AGN disks are observed to have elevated metallicities across redshift \citep{Dors+14,Onoue+2020}.
Muffled SNe may be a contributing source by infusing the disk with localized pockets of metals.
SNe in the breakout region will similarly enhance the midplane metallicity, but the plume that emerges from the disk will neither expand into the steady-state, uniform medium as we have modeled nor retain the quasi-hemispheric shape we see in Fig.~\ref{fig:timeseries-composite-xz}.
Rather, it will encounter an intense radiation field emitted by the hot corona of the BH that drives outflows along the disk surface, blowing the plume to larger radii \citep{CrenshawKraemerGeorge03,ProgaKallman04}.
As the wind drives the SN ejecta at speeds of $10^2-10^5$ km s$^{-1}$ \citep{Parker+17,Mehdipour+17}, it will spiral outward, scattering nucleosynthetically enriched material across the outer disk.
If SNe are a significant contributor to AGN metal fractions, that likely indicates a top-heavy stellar mass function \citep{FanWu23} and that embedded stars evolve off the main sequence during the AGN's active phase \citep{HuangLinShields23} as opposed to doing so after the disk fully accretes \citep{Cantiello21}.

The SG disk model may only represent one phase of the accretion process over the course of an AGN lifetime.
Disks in lower states of accretion will be more tenuous like the TQM model, allowing the SN to expand further into the midplane, excavate a larger portion of the disk atmosphere, and cause more disruption.
Neither of these models consider magnetic fields that are carried into the disk by parcels of gas as they accrete from the torus.
As magnetic fields build, the additional pressure becomes an important component of the overall disk pressure.
Even fields that produce a plasma $\beta \equiv P_{\rm gas} / P_{\rm mag} = 10^{2.5}$ can elevate $H$ relative to disks with much weaker fields at $\beta=10^4$ \citep{Gerling-Dunsmore+25}.
Since $H$ is the primary characteristic of an AGN disk for determining shock breakout, the magnetic fields could move the muffling radius inwards if the gas density remains similar to cases with no magnetic fields.
In extreme cases, disk models following gas flows in cosmological simulations down to within a few Schwarzschild radii demonstrate buildup of strong magnetic fields reaching $\beta \sim 10^{-6} - 10^{-2} \ll 1$ \citep{Hopkins+25}.
These flux-frozen disks suppress star formation entirely, implying the progenitors of AGN-SNe must be captured by the disk from the nuclear star cluster \citep{Fabj20,Nasim+23}.

In the case where star formation can occur in the disk, the AGN-SN rate is estimated to be $\dot{N}_{\rm SN}\sim10^{-2}\,{\rm yr}^{-1}$ for $\Mbh=10^8\,\Msun$ \citep{HuangLinShields23}.
If the density of AGN in the Universe is $n_{\rm AGN}=10^6$ Gpc$^{-3}$ \citep{OsterbrockFerland06}, the rate density of AGN-SNe is $\mathcal{R_{\rm SN}} \sim 10^4$ yr$^{-1}$ Gpc$^{-3}$.
Phenomena related to AGN activity may be visible for a range timescales spanning $10^4$ \citep{Morganti17} to $10^6$--$10^8$ yr \citep{Biava+21} depending on the specific mechanism being probed, but some modeling suggests the dense disks we consider here have a lifetime on the order of $\tau_{\rm AGN}=10^6$ yr \citep{Angles-Alcazar+21}.
Given $\dot{N}_{\rm SN}$ over that lifetime, the number of SNe that occur within the disk for a typical AGN is $\mathcal{N}_{\rm SN}=10^4$ AGN$^{-1}$.

However, not all SNe will break free from the disk as we have shown.
If we assume SNe occur evenly throughout a disk with outer radius of $10^7\,\Rs$ and a muffling radius at $\Rb\sim10^4\,\Rs$, then we need to scale by the fraction of the disk area that allows breakout $f_{\rm b} =  (R_b / R_{\rm disk})^{2} = 10^{-6}$.
Thus, the breakout rate per AGN becomes
\begin{equation}
    \mathcal{N_{\rm b,SN}} \sim 10^{-2} \left( \frac{f_{\rm b}}{10^{-6}} \right) \left( \frac{\dot{N}_{\rm SN}}{\rm 10^{-2}\ \frac{SN}{yr\ AGN}} \right) \left( \frac{\tau_{\rm AGN}}{\rm 10^6\ yr} \right) \mathrm{AGN^{-1}},
\end{equation}
and the volume rate similarly reduces to
\begin{eqnarray}
	\mathcal{R_{\rm b,SN}} &\sim& 10^{-2} \left( \frac{f_{\rm b}}{10^{-6}} \right) \left( \frac{\dot{N}_{\rm SN}}{\rm 10^{-2}\ \frac{SN}{yr\ AGN}} \right) \nonumber \\ 
	&& \times \left( \frac{n_{\rm AGN}}{\rm 10^6\ \frac{AGN}{Gpc^3}} \right) \mathrm{Gpc^{-3}\ yr^{-1}}
\end{eqnarray}

This study explored the possibility of SN shocks breaking free from the disk.
\citet{Moranchel+21} and \citet{Grishin+21} approached this question with different methods yet achieved similar results for midplane and off-set scenarios.
Whether radiation emitted by a SNe can be observed against the bright AGN disk is a separate matter and depends on the optical depth through the disk to the detonation site.
For example, even though the shock in the model placed at $R=10^5\,\Rs$ was muffled, the SG disk model is optically thin at that radius when $\Mbh=10^8\,\Msun$, and photons directly emitted a SN may still escape.
We will present detailed post-processing radiative transfer calculations to derive light curves for our shearing box simulations in a companion paper.

As large surveys like the Legacy Survey of Space and Time (LSST) generate baselines for substantially more AGN \citep{Kovacevic+22}, if we are able to constrain the rates of observable and non-observable AGN-SN, we will also have constraints on AGN disk properties such has their sizes and densities.
If the rate is small, then either most SNe are muffled in a puffed-up, low-density AGN disk or the disk is typically small.
In contrast, a large rate suggests AGN disks are likely large, geometrically thin, dense, and thus able to capture or produce stars at a high rate.

Though we have considered Type II SNe ejecting $10\,\msun$ envelopes, AGN-embedded stars may not evolve similarly to stars in the field, changing our rate estimates and details of the explosions such as the ejecta mass and energy of the SN \citep{Cantiello21, DittmannJermynCantiello23}.
Similarly, Type Ia novae from accreting white dwarfs and kilonovae from binary neutron stars formed by the disk likely evolve differently when embedded in an AGN disk.


\section{Conclusions}

Massive stars in nuclear star clusters on orbits embedded in AGN accretion disks may detonate as SN from within the disk.
The shock can break free from the disk if its internal pressure exceeds the ambient disk pressure and is muffled otherwise.
The exact outcome depends on the radial and vertical location of the detonation as described by Eq.\,(\ref{eqn:C-parameterized}).
We performed hydrodynamic three-dimensional shearing box simulations at two locations on either side of the muffling boundary of embedded SNe and found agreement with the analytic prediction for shock breakout from an SG-type AGN disk orbiting a black hole of mass $\Mbh=10^8\,\Msun$.
Specifically, we found:

\begin{itemize}
    \item In general, SNe break free from the inner disk, whereas the outskirts of AGN disks muffle them.
    \item In SG disks, when $\mbh=10^6\,\msun$, the muffling boundary occurs at $R_{\rm m}=5\times10^6\,\rs$ ($\sim10^6$ AU) and moves inward to $R_{\rm m}=100\,\Rs$ ($\sim100$ AU) in disks around $\mbh=10^9\,\msun$.
    \item In TQM disks, the muffling radius lies at $R_{\rm m}=10^6\,\Rs$, regardless of the SMBH's mass.
    \item The muffling radius for SNe that occur at $z>1\,H$ shifts outwards by approximately one order of magnitude in comparison with midplane detonations.
    \item When SNe detonate away from the midplane the ejecta traveling to the far side of the disk may remain shocked or emerge as subsonic flows depending on the gas density and scale height, where the optical thickness will determine whether photons do the same.
    \item When the SN occurs away from the midplane, the shock propagating away from the disk expands rapidly in comparison to the shock propagating towards (and possibly through) the midplane, slowing down as it climbs the density and pressure gradients.
    \item Considering a lifetime of $\tau_{\rm AGN}=1$ Myr and a SN rate in the disk of $\dot{N}_{\rm SN} \sim 10^{-2}$, we estimate a rate of $\mathcal{R}_{\rm b} \sim 10$ SN AGN$^{-1}$ yr$^{-1}$ breaking free from the disk and estimate a volumetric rate density of $\mathcal{R}_{\rm SN} \sim 10$ Gpc$^{-3}$ yr$^{-1}$ for the local Universe.
\end{itemize}

As Vera Rubin Observatory and LSST expand our transient catalogs, automated search algorithms may identify light curves with previously-unobserved characteristics or features.
One potential source of these new transients could be AGN-embedded SNe whose light curves are altered by an AGN disk.
In a companion paper we will describe results from our radiative transfer analysis of the models presented here.
Combining radiative and kinematic SNe muffling may reveal information about an average disk's structure, leading to a broader understanding of AGN accretion physics and galactic nuclei in general.

\begin{acknowledgments}
We thank the anonymous referee for a thorough report that greatly improve the paper.
HEC, KESF, and BM are supported by NSF AST-2206096.
M-MML is partly supported by NSF grants AST18-15461 and AST23-07950.
KESF and BM are also supported by NSF AST-1831415 and Simons Foundation Grant 533845.

The simulations were performed on Stampede at NSF’s Texas Advanced Computing Center (TACC) using XSEDE/ACCESS grant TG-AST140014, and the Discovery cluster at New Mexico State University \citep{TrecakovVonWolff21}. This work utilized resources from the New Mexico State University High Performance Computing Group, which is directly supported by the National Science Foundation (OAC-2019000), the Student Technology Advisory Committee, and New Mexico State University and benefits from inclusion in various grants (DoD ARO-W911NF1810454; NSF EPSCoR OIA-1757207; Partnership for the Advancement of Cancer Research, supported in part by NCI grants U54 CA132383 NMSU).
\end{acknowledgments}

\software{\texttt{Astropy} \citep{Astropy+13,Astropy+18,Astropy+22}, \texttt{Athena} \citep{Stone+08}, \citep{Hunter07-Matplotlib}, \texttt{NumPy} \citep{Harris+20-Numpy}, \texttt{pAGN} \citep{Gangardt+24}, \texttt{SciPy} \citep{Virtanen+20-Scipy}, \texttt{jupyter} \citep{Jupyter}, \texttt{IPython}\citep{IPython}, \texttt{matplotlib} \citep{matplotlib1, matplotlib2}.}

\bibliography{master}{}
\bibliographystyle{aasjournalv7}

\end{document}

%% file: preamble.tex
\usepackage{amsmath,amssymb,empheq}
\usepackage{nicefrac,xfrac}
\usepackage[normalem]{ulem}
\usepackage{color}
\usepackage{cancel}
\usepackage{graphicx,subcaption,lipsum}

\usepackage{outlines}
\usepackage{enumitem}
\setenumerate[1]{label=\Roman*.}
\setenumerate[2]{label=\Alph*.}
\setenumerate[3]{label=\roman*.}
\setenumerate[4]{label=\alph*.}
\setlist{nosep}

\newcommand{\msun}{M_{\odot}}
\newcommand{\Msun}{\msun}
\newcommand{\Mej}{M_{\rm ej}}
\newcommand{\rhoej}{\rho_{\rm ej}}
\newcommand{\Mbh}{M_{\rm BH}}
\newcommand{\mbh}{\Mbh}
\newcommand{\gcmcube}{\rm{g\ cm^{-3}}}

\newcommand{\cs}{c_{\rm s}}

\newcommand{\mProt}{m_{\rm p}}
\newcommand{\Rs}{R_{\rm s}}
\newcommand{\rs}{\Rs}

\newcommand{\Rsn}{R_{\rm SN}}
\newcommand{\Esn}{\mathbb{E}_{\rm SN}}
\newcommand{\Rb}{R_{\rm b}}
\newcommand{\esn}{\varepsilon_{\rm SN}}

\newcommand{\tpds}{t_{\rm PDS}}
\newcommand{\Rpds}{R_{\rm PDS}}
\newcommand{\fkinetic}{f_{\rm K}}

\renewcommand{\vec}[1]{{\boldsymbol{#1}}} 
\newcommand{\ptderiv}[1]{\frac{\partial{#1}}{\partial{t}}}
\newcommand{\del}{\vec{\nabla}}

\newcommand{\Div}{\del\cdot}

\definecolor{brown}{rgb}{0.42,0.24,0.07}
\definecolor{darkgreen}{rgb}{0.0,0.6,0.00}
\definecolor{purple}{rgb}{0.7,0.0,0.7}
\definecolor{black}{rgb}{0.0,0.0,0.0}

\newcommand{\eq}[1]{Eq.~(\ref{#1})}

\newcommand{\C}{\mathcal{C}}